\documentclass[journal=langmd5, manuscript=article]{achemso}

\usepackage{hyperref}

\usepackage[T1]{fontenc}
\usepackage{amsmath, amssymb, amsfonts}
\usepackage{multirow}
\usepackage{bm ,textcomp}
\usepackage{float}
\usepackage{graphicx}
\graphicspath{{./figures/}}

\expandafter\ifx\csname package@font\endcsname\relax\else
 \expandafter\expandafter
 \expandafter\usepackage
 \expandafter\expandafter
 \expandafter{\csname package@font\endcsname}%
\fi
\title {Three-dimensional modelling of polygonal ridges in salt playas}

\author{R. A. I. Haque}
\affiliation[sxc]{Physics Department, St. Xavier's College, Kolkata 700016, India}
\alsoaffiliation[cmprc]{Condensed Matter Physics Research Centre, Jadavpur University, India}
\author{A. J. Mitra}
\affiliation[montana]{Mathematical Sciences, Montana Tech, Butte, MT 59701, United States}
\author{T. Dutta}
\affiliation[sxc]{Physics Department, St. Xavier's College, Kolkata 700016, India}
\alsoaffiliation[cmprc]{Condensed Matter Physics Research Centre, Jadavpur University, India}
\email{tapati\_dutta@sxccal.edu}

\date{\today}%

\begin{document}

\begin{abstract}
Salt playas with their tessellated surface of polygonal salt ridges are beautiful and intriguing, but the scientific community lacks a realistic and physically meaningful model that thoroughly explains their formation. In this work, we investigated the formation phenomena via suitable three-dimensional modelling and simulation of the dynamical processes that are responsible. We employed fracture mechanics, principles of energy minimization, fluid and mass transport in fracture channels and processes of crystallization and self organisation to finally replicate the almost Voronoidal pattern of salt ridges that tessellate salt playas. The model is applicable to playas having different salt compositions, as the effect of the salt diffusion coefficient and critical salinity at supersaturation for a particular ambient condition are factored in. The model closely reproduces the height distribution and geometry of the salt ridges reported in the literature. Further, we prove that the final stable polygonal geometry of the salt playas is an effort towards the total minimization of system energy.
\end{abstract}

\maketitle



\section{Introduction}
Salt playas or salt deserts often display beautiful patterns of salt ridges that tessellate the entire saline pan into polygons most of which are pentagons and hexagons.  Saline pans are flat shallow depressions covered with layers of salt that are normally dry, except when flooding or heavy rain turns the pan into a temporary lake. Over a long period of time, the pans go through repeated wetting and drying cycles\cite{Lowenstein1985}. The salt ridges develop annually up to a few centimetres on the top surface, can cover an area ranging from 1 $km^2$ to a few thousand $km^2$, and overlie salt-water saturated thin-bedded clays and gypsum deposits\cite{Christiansen1963, Hardie1978book, Rettig1980}. Over the last few decades, researchers have tried to explain the phenomena of large polygonal salt ridges that form a mosaic on salt playas by using fracture\cite{Krinsley1970} or buckling models\cite{Christiansen1963} of the surface layer, or by modelling complex fluid dynamics in the sub-surface of these regions\cite{Wooding1960, Homsy1976, Wooding1997, Wooding2007, Dam2009, Lasser2023}.  However, these models do not offer any explanation that justifies the images of the rough pentagonal or hexagonal Voronoi-like tessellation of several square kilometres of the salt basins by salt ridges.

\begin{figure}[h]
\centering
\includegraphics[width=0.85\linewidth]{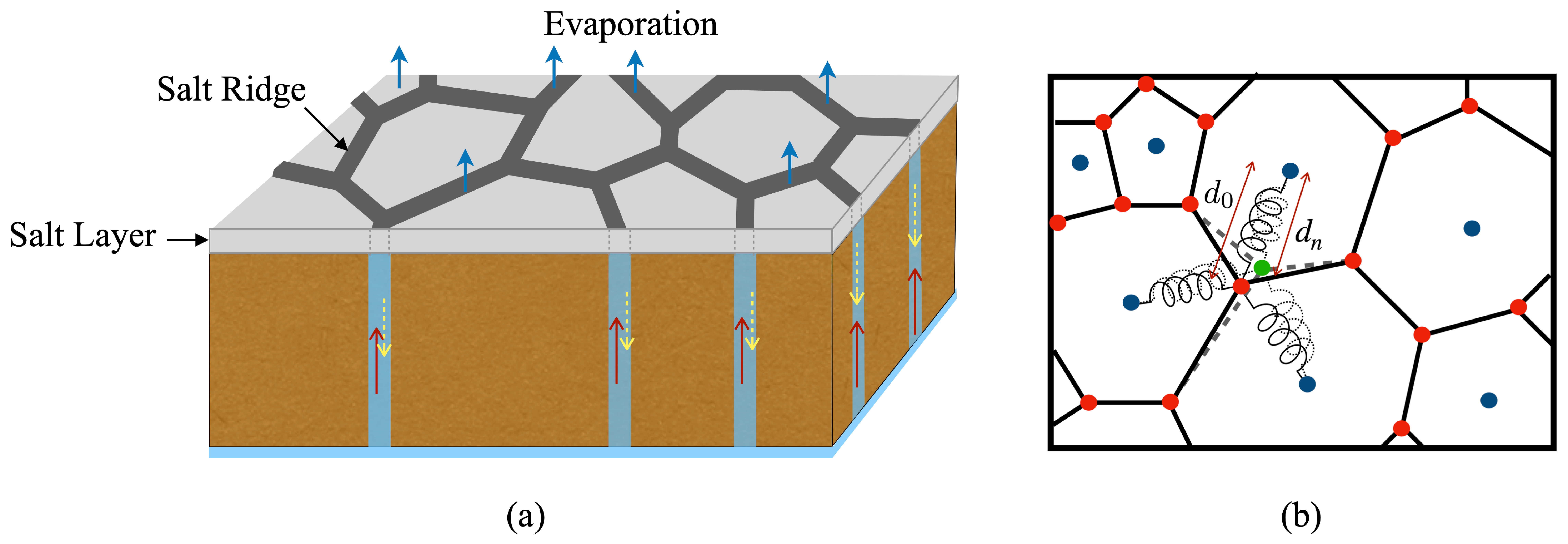}
\caption{(a) Schematic diagram of the salt playa. The grey layer represents a salt layer on top of the clay surface. Polygonal salt ridges develop along the boundaries of the polygonal cracks as saline water gets transported from the subsurface water table along vertical crack channels. Directions of advection and diffusion are indicated by red and white arrows respectively. The evaporation flux direction is indicated by blue arrows. (b) Spring model to simulate desiccation cracking. The crack ped vertices are marked by red dots. The ped centroids are marked by blue dots. $d_0$, the natural spring length between the centroid (blue) and vertex (red) indicated by an arrow.
A red vertex is acted upon by the elastic forces, represented by springs, from its surrounding polygons. The vertex is displaced to a new position (green dot) after $n$ desiccation steps, where it remains connected to the polygon centroids with springs
(shown by dotted springs) of new lengths $d_{n}$ (indicated by arrow).  }
\label{fig:schematic}
\end{figure}

In this paper, we provide an explanation of the polygonal salt patterns by incorporating principles of fracture mechanics, fluid dynamics, mass transport and finally the principles of crystallization. During dry seasons, desiccation of the surface clay layers fracture to release accumulated stress. A growing crack stops once it meets another crack line, in the process tessellating the top surface into polygonal crack mosaics that resemble Gilbert tessellation\cite{Gilbert1967}. With time, the stress propagates to the lower layers and the crack network extends vertically downwards. The crack channels that extend from the surface to the subsurface water table act as primary water-conducting channels. A schematic diagram is shown in Figure \ref{fig:schematic}a.

During desiccation water evaporates from the surface layer and saline water from the subsurface water table (or trapped in the interstitial pores of the clay region) is pulled up due to advection caused by evaporation flux. As water evaporates it precipitates the dissolved salt thereby increasing the salt concentration of the top layer of the channels. A concentration gradient of salt can develop in the conduits where the salt concentration of the subsurface water table is assumed to have a constant lower value. This initiates a diffusion process from the top layer to the bottom of the channel as indicated in Figure \ref{fig:schematic}a. The mass transport of salt is determined by the combined processes of advection and diffusion and defined in terms of the Peclet number $Pe=UL/D$, where $U$ is the advection velocity, $L$ is a characteristic length scale and $D$ is the diffusion coefficient of the solution. If the competition between advection and diffusion causes the top layers of the crack channels to become supersaturated, crystallization of salt is facilitated along the crack edges.

Let the equilibrium salinity of the water table be $c_0$, and the critical salt concentration at supersaturation be $c_{cr}$; then the difference in salt concentration $\Delta c = c_{cr} - c_{0}$ determines the rate of salt crystallization along the surface crack edges for a given ambient condition. Salt crystals keep accumulating along the crack edges to form ridges which self-organise to stable structures governed by the angle of repose of salt mounds. The pattern of the ridges follows the pattern of the crack network on the clay surface. Over a long period of time, as the clay surface goes through multiple wetting-drying cycles,  the pattern of the crack network changes to attain the minimum energy configuration. The angles created at the junctions where two or more crack lines meet, evolve from $90^\circ$ to $120^\circ$\cite{Goehring2010} and the crack pattern and therefore the salt ridge pattern becomes Voronoi-like\cite{Ruhul2023chaos}.

The Voronoi diagram is a collection of regions that tessellates a plane around a given set of points (seeds) such that every point in a region is closest to the corresponding seed. Given a finite number of points $x_{1},x_{2} \dots x_{n}$ on a subset $\Omega$ of the plane, the Voronoi regions are defined as $\Omega_{i} = \{x \in \Omega : ||x-x_{i}|| \leqslant ||x-x_{j}|| \text{ for all } j \neq i\}$. If the centroids of the polygons coincide with the seeds, then the Voronoi diagram is known as the centroidal Voronoi diagram. Many Voronoi-like patterns exist in nature and are often found to be associated with the optimization of material or energy\cite{Budkewitsch1994, Jagla2002, Ruhul2022, Ruhul2023chaos}. Images of salt ridges tessellating salt playas as reported by naturalists appear to resemble Voronoi mosaics.  
The Voronoi-ness of a real mosaic can be measured by comparing it with the Voronoi diagram generated using the centroids of the polygons as seeds. The Hausdorff metric\footnote{For compact subsets $A,B \subset \mathbb{R}^2$, the Hausdorff Metric $d_H(A,B)$ is defined as $\displaystyle{d_H(A,B)= \max\{\max_{a \in A}d(a,B),\max_{b \in B}d(b,A)\}}$, where $d(x,C)= \inf \{\| x-c \| : c \in C \}$.} is used to measure the deviation of the salt ridge patterns from the corresponding Voronoi diagram.

In the following sections, we shall describe the model in detail starting from crack development under wetting drying cycles, to the governing equations of fluid flow in the conduits, to the process of crystallization and finally the stable formation of salt ridges. This will be followed by discussion of results of the simulation and the final conclusions.

\section{The simulation model}
\subsection*{Initial cracks}
As water evaporates from clay or soil surfaces during desiccation, the matrix (clay/soil) develops stress due to capillary pressure in the pores \cite{Goehring2015_book}.  Desiccation causes the surface clay layers to fracture, releasing accumulated stress. Thus all crack faces are regions of zero stress. In our simulation, the cracks begin from random points and grow along a straight line till they collide with an existing crack. The initial crack network of the clay surface is mimicked by a Gilbert tessellation \cite{Gilbert1967, Anamika2022} generated from a homogeneous Poisson point distribution of seeds. With desiccation, the lower layers of clay develop stress due to evaporation of water from the interstitial pores and the cracks extend vertically downwards with every drying step \cite{Ruhul2023chaos}. These vertically extended cracks act as conduits that eventually connect with the subsurface water table. 
\subsection*{Wetting-drying}
The effect of wetting-drying is modelled via a connected spring network which is often used to mimic fracture mechanics\cite{Bolander2005, Khatun2012, Kitsunezaki2013, Tarafdar2019, Sadhukhan2019}. After generating the Gilbert tessellation, all the polygons and nodes are detected. The nodes are connected to the centroids of those polygons that share the node via springs with spring constant $k$. The natural length of a spring is set to $d_0$ - the distance between the node and the centroid of the polygon that the spring is connected to. During desiccation, the length of each spring contracts and during wetting the length expands according to the following rule --
\begin{equation}
    d_n = d_0(1 \mp \frac{q}{r^n})
    \label{w-d rule}
\end{equation}
Where $q$, $r$ are constants and the negative sign is for drying and the positive sign is for wetting. It is assumed that while the drying process causes the shrinking of clay layers, during wetting the layers relax a little as water fills the interstitial pores space causing the clay to swell to some extent. The spring length is $d_n$ after $n$ successive drying and wetting steps; one time step is made up of a single drying followed by a  wetting process, a schematic representation of the spring-model is shown in Figure \ref{fig:schematic}b.  The wetting-drying rule given by eq \ref{w-d rule} is guided by experiments on clay evaporation rate under ambient conditions where complete drying is equivalent to $\approx 63\%$ of weight loss \cite{Sadhukhan2007}. The parameters $a$ and $b$ represent ambient conditions and clay characteristics respectively. The position of a node that is connected to the centroids of its neighbouring polygons with springs, is changed according to --
\begin{equation}
    \Delta \mathbf{x} = \sum_{i}\Delta \mathbf{d}_{i}
\end{equation}
where the summation is over the springs that are connected to the node and $\Delta \mathbf{d}_{i}$ is the change in length of the $i^{th}$ spring. Here $\Delta \mathbf{d}_{i} = (d_{n} - d_{0})\hat{d}_{i}$ with $\hat{d}_{i}$ being the unit vector directed from the node toward the centroid of its $i^{th}$ neighbouring polygon. The crack segments follow the movements of the nodes and the shape of the crack network evolves with wetting-drying cycles.\\
\subsection*{Fluid dynamics and crystallization}
During desiccation water evaporates from the top surface of the salt layer causing a suction pressure that draws water from the subsurface water table to replace the evaporated water. The desiccation pull is represented by a pressure $p$ which in effect controls the water flow through the channels. Subsurface water, specially in salt playa regions, is high in salinity. Thus as water evaporates from the water-carrying channels, the dissolved salt gets deposited on the channel's top layers, increasing the salinity here.  A salt concentration gradient develops that is directed from the upper to the lower layers of the channel, thereby initiating a diffusion flow in this direction, Figure \ref{fig:schematic}b. The competition between advection and diffusion flows, characterised by the Peclet number $Pe$, controls the salinity profile of the channels.

The governing equations that are solved following a finite difference method are 
the continuity equation for an incompressible fluid, the Navier-Stokes equation to solve the pressure and velocity field in the channel and the advection-diffusion equation:
\begin{eqnarray}
    \nabla \cdot \mathbf{v} &=& 0 \\
    \frac{\partial \mathbf{v}}{\partial t} &=& -\frac{1}{\rho} \nabla p + \eta \nabla^{2}\mathbf{v} \\
    \frac{\partial c}{\partial t} &=& D\nabla^{2}c - \mathbf{v} \cdot \nabla c
\end{eqnarray}
where $\mathbf{v}$ is solution velocity in a given grid cell, $\rho$ is the density, $\eta$ is viscosity, $c$ is salinity of water at any instant of time and $D$ is diffusion constant. The details of the discretisation are found in other works of the authors \cite{Sadhukhan2018, MDC2015}. 

The evaporation flux is proportional to the average vertical component $U$ of the velocity of water on the top exposed surface of any channel due to the pressure $p$. Therefore the number of water cells evaporating per unit area at each time step $\delta t$, i.e. the evaporation flux $Q$ is given by\\
$$Q = \frac{\phi U \delta t}{\delta x^{3}}$$\\
 where $\delta x^{3}$ is the volume of a single grid cube of length $\delta x$, and $\phi$ is the porosity of the top surface of the salt playa. The probability of water evaporating from the top surface of a channel is guided by the evaporation flux, and done stochastically using a random number generator. When a cell's water evaporates, the salt contained in the water is equally distributed among the neighbouring cells.
At the top layer of the channels, if the salinity of water near the channel wall regions reaches a salinity $\geqslant c_{cr}$, say $c_{cr}+ \delta c$, the excess salt $\delta c$ crystallizes and deposits on top of the channel wall. The irregularities of the channel walls act as seeds for the crystallization process. It is assumed that the latent heat of crystallization released during crystallization does not affect the water temperature since water has a high latent heat. Examination of Figure \ref{fig:3D-pattern}b shows that the salinity gradient diffuses quickly to equilibrium within a few top layers of the pore channel. Hence solution density is kept constant in the model.

With wetting-drying cycles of the playas over many years, the crack mosaics evolve to form a stable equilibrium geometry that is referred to as the `matured' state. Salt deposits along the crack edges shift with shifting crack pattern until the crack mosaic reaches maturation. Upon maturation when the crack pattern becomes static, salt crystals accumulate along the cracked boundaries and develop into ridges. In our model, the ridges are allowed to form stable structures through self-organisation by distributing the deposited crystals to nearest neighbour sites that have a lower potential energy (lower height). The self-organisation of the crystals occurs while maintaining an angle of repose of $63^{\circ}$. An experimental determination of table salt (NaCl) procured from Tata Chemicals Ltd. showed an angle of repose of $58.5^{\circ}$, Figure \ref{fig:3D-pattern}c. The angle of repose is dependent not only on the chemical composition of a salt, but also on parameters like grain size, shape, purity and its hygroscopic properties. Our model being generic in nature, we have maintained the angle of repose to $\approx 63^{\circ}$ for each of simulation.

\begin{figure}[H]
\centering
\includegraphics[width=0.95\linewidth]{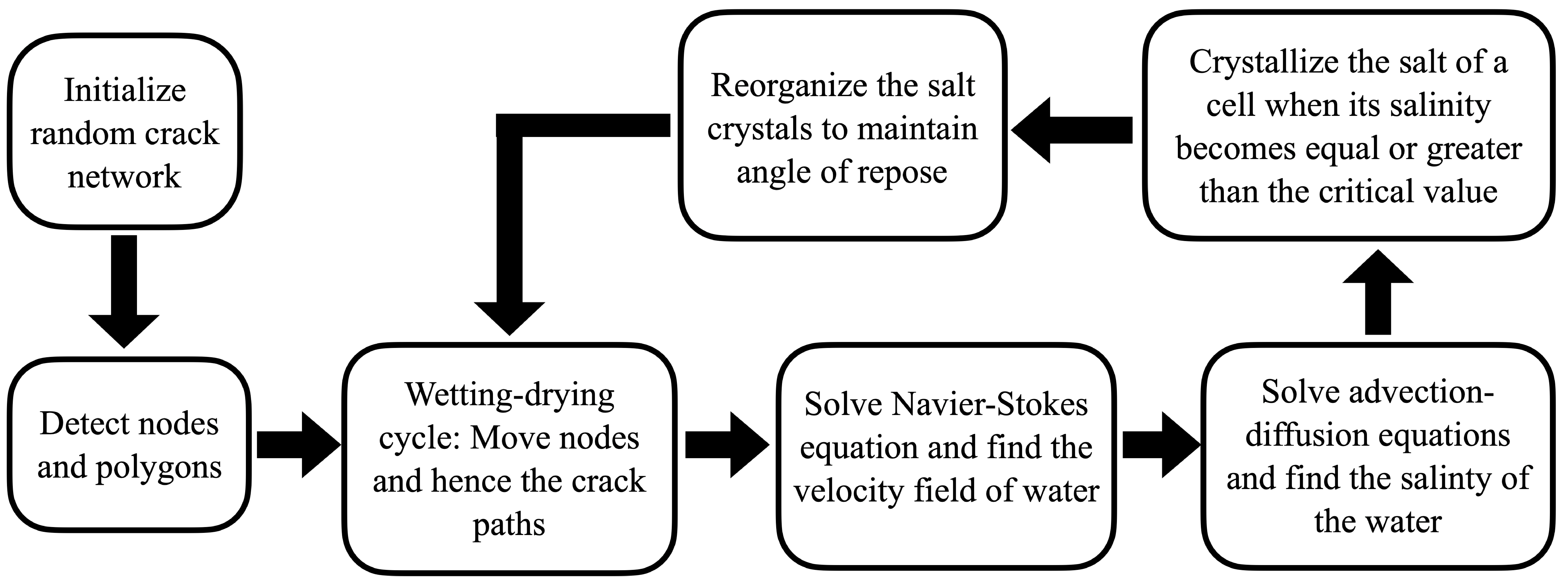}
\caption{Flowchart of simulation.}
\label{flowchart}
\end{figure}

The flowchart of the simulation starting from surface crack formation and its evolution in time, control of evaporation flux that provides the draw pressure of saline water from the subsurface water table to the surface - the determination of pressure and velocity field in the crack channels - distribution of salt concentration in the channels due to the combined effect of advection and diffusion - crystallization of salt on the walls - and finally the self-organisation of the growing salt ridge to stable structures, is displayed in Figure \ref{flowchart}.

\section{Results and Discussion}
\begin{figure}[H]
\centering
\includegraphics[width=0.75\linewidth]{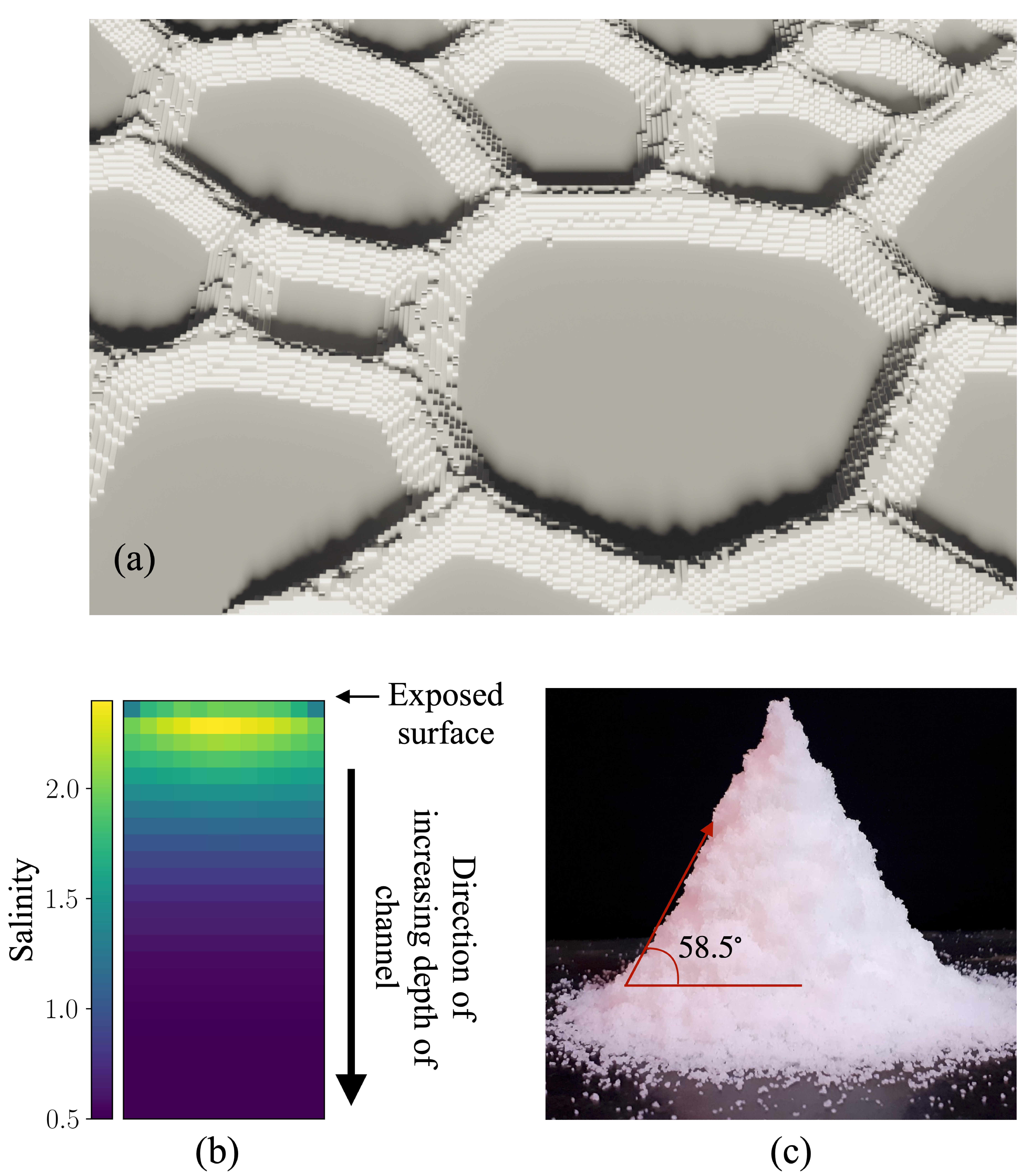}
\caption{All parameters are expressed in suitable units. (a) Simulated salt ridge patterns in 3-D. (b) Salinity profile in a typical pore channel. Salinity concentration in the channel is indicated by the colour legend, deep blue corresponds to the initial salinity $c_0$ while the yellow colour corresponds to the highest value of salinity. (c) Experimental measurement of angle of repose of wet table salt (Tata Chemicals Ltd.) shows $58.5^{\circ}$}
\label{fig:3D-pattern}
\end{figure}

The 3-D simulated polygonal salt ridges tessellating a playa as obtained from our model is displayed in  Figure \ref{fig:3D-pattern}a. The lateral striations observed along the ridge walls are indicative of the self-organising of salt deposits to maintain the angle of repose to $63^{\circ}$.

 The rate of salt crystallization along the crack edges is a resultant of the combined effects of evaporation flux that determines the suction pressure $p$, the diffusion coefficient $D$ of the saline solution, the equilibrium value of salt concentration of the subsurface water table $c_{0}$ and the critical salt concentration $c_{cr}$ at which the solution becomes supersaturated and crystallization is initiated. The aforesaid parameters are in the CGS units in our simulation. For any salt solution, $D$ is a constant. An initial examination further revealed that the concentration difference $\Delta c$ rather than $c_{0}$ or $c_{cr}$ is the important factor that determines the start of crystallization $t_{cr}$.  Therefore we shall display and discuss our results with respect to variation of $\Delta c$. Figure \ref{fig:3D-pattern}b is a screenshot of a typical distribution of salinity in a pore channel under the combined effects of $p, D$ and $\Delta c$ during the evaporation process. The colour legend moves from dark blue to yellow as the salinity increases from the initial concentration of $c_0$ to a high concentration of $c_{cr}$. The top layer of the channel indicates a lower salinity than the layer just below as it has contributed towards crystallization on the crack edge. The blue colour adjacent to the wall on the topmost layer has deposited the salt towards the formation of crystals and hence shows a lower salinity.

   Though it is almost customary to describe advection-diffusion flow processes through variation of the Peclet number $Pe$, we observe that the effect of advection and that of diffusion have very marked differences on the first crystallization time $t_{f}$ which gets masked if their combined effect through $Pe$ is used as a parameter. Hence we present and discuss the simulation results with respect to $p$ (responsible for the advection velocity $U$) and $D$ separately. We assume that the evaporation flux is vertical to the surface plane and not affected by other perturbations like wind velocity or changes in temperature or humidity.\\
 \textbf{First crystallization time $t_{cr}$ dependence on evaporation flux}\\
\begin{figure}[H]
\centering
\includegraphics[width=0.85\linewidth]{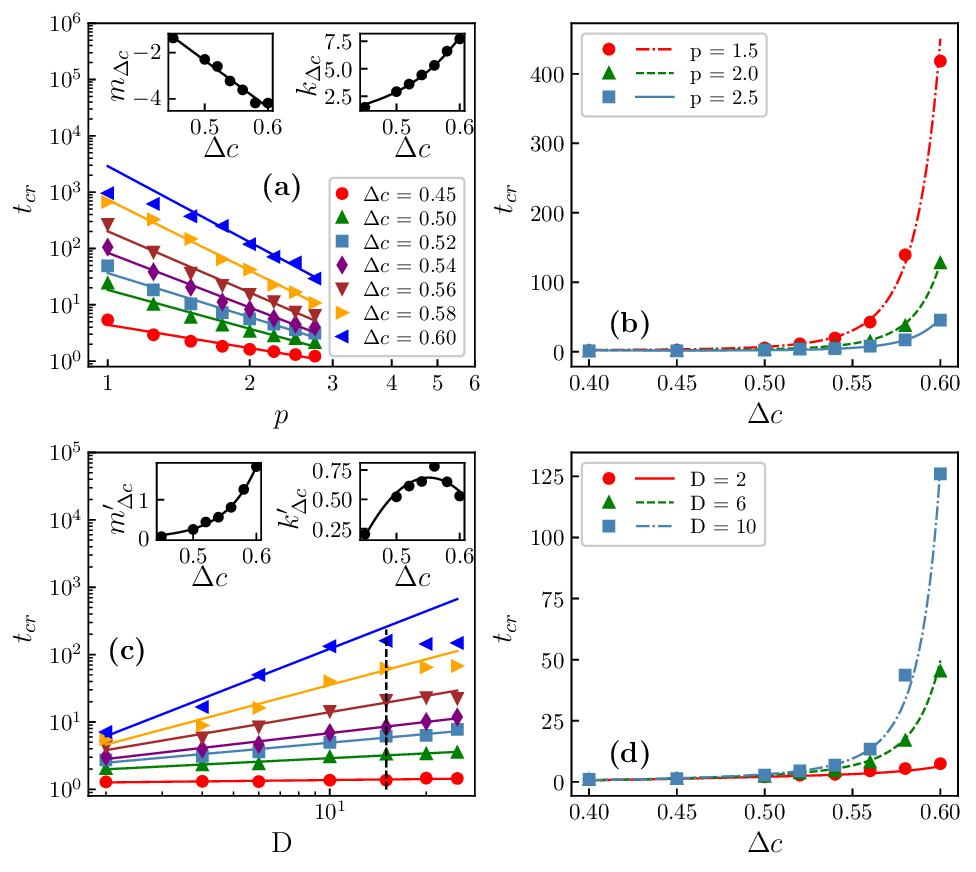}
\caption{All parameters are expressed in suitable units. (a) Log-log plot of variation $t_{cr}$ with  $p$ for different $\Delta c$. Insets show the variation of scale exponent $m_{\Delta c}$ and $k_{\Delta c}$ with $\Delta c$. (b) Validation of analytical relation between $t_{cr}$ and $\Delta c$. The symbols represent simulated data points and the curves are obtained from eq \ref{t-deltac}. (c) Log-Log plot of $t_{cr}$ versus $D$ for various $\Delta c$. Insets show the variation of scale exponent $m^{\prime}_{\Delta c}$ and $k^{\prime}_{\Delta c}$ with $\Delta c$. The colour legend is identical to (a). (d) Validation of analytical relation between $t_{cr}$ and $\Delta c$. The symbols represent simulated data points and the curves are from eq \ref{t-deltac-prime}.}
\label{t-c}
\end{figure} 
 
In the study of the dependence of crystallization on suction pressure $p$ for variation in  $\Delta c$, the diffusion coefficient is held constant at $D=6$ which though chosen heuristically, is guided by our results discussed in the forthcoming sections. We shall note that at higher $D$ values, $t_{cr}$ increases exponentially while too small a value is almost masked by the effect of advection. A log-log plot of $t_{cr}$ versus $p$, Figure \ref{t-c}a, shows a power-law dependence with exponent $m_{\Delta c}$ and slope $k_{\Delta c}$, both being functions of $\Delta_c$. The unit of time is taken to be equal to one time step of our simulation.
\begin{equation}
\log{t_{cr}} = m_{\Delta c}\log{p} + k_{\Delta c}
\label{m}
\end{equation}
 The first crystallization time $t_{cr}$ increases in an almost linear manner for low evaporation rates and for $\Delta c \leqslant 0.56$; however $t_{cr}$  increases non-linearly thereafter. When  $\Delta c$ is small,  even a small evaporation flux $p$ induces sufficient advection to enable crystallization in the top layer of the channels, hence $t_{cr}$ is attained quickly. The effect of diffusion is felt only when $\Delta c$ is sufficiently large as diffusion flow occurs in opposition to advection, see Figure \ref{fig:schematic}. Hence a longer time is required before the top layer of the channels reaches the state of super-saturation. However, if advection is increased sufficiently by increasing evaporation flux, the effect of diffusion is almost negated by strong advection and the value of $t_{cr}$ asymptotically tends to a constant value for any $\Delta c$. The insets of Figure \ref{t-c}a show that $m_{\Delta c}$ decreases linearly for the entire range of $\Delta c$ and $p$ except for $\Delta c = 0.60$ at $p\leq 1.8$. $k_{\Delta c}$ increases exponentially with $\Delta c$, i.e.
\begin{eqnarray}
 m_{\Delta c} &=& \alpha \Delta c +\beta \\
 k_{\Delta c} &=& a e^{b \Delta c}
\label{m-deltac}
\end{eqnarray}
 where, $\alpha$, $\beta$, $a$ and $b$ are functions of $D$ and hence, constants for a given $D$. Figure \ref{t-c}a shows a negative $\alpha$ with $\beta$ positive but $k_{\Delta c}$ shows a positive exponential increase with $\Delta c$.
 Combining eqs (\ref{m} -- \ref{m-deltac}), 
\begin{equation}
 \log {t_{cr}}= (\alpha\Delta c +\beta)\log p + a e^{b \Delta c}
\label{t-p-deltac}
\end{equation}
i.e.,\\
\begin{equation}
 t_{cr}= \exp\left[{(\alpha\Delta c +\beta)\log p + a e^{b \Delta c} }\right]
\label{t-deltac}
\end{equation}
Equation \ref{t-p-deltac} captures the relation between the time of first crystallization $t_{cr}$, $p$ and $\Delta c$.
In eq \ref{t-deltac}, the first term of the exponential is negative as $\alpha$ is negative, and the second term is positive as both $a$ and $b$ are positive. For the range of $p$ studied here, the positive term is always larger than the first term and this is reflected in the positive exponent of the exponential. For large $p$, $t_{cr}$ tends to $1$ as the exponent $\longrightarrow 0$. Figure \ref{t-c}b displays simulation results for the variation of $t_{cr}$ and $\Delta c$ for different $p$ and constant $D$ values; the lines in the graph are plotted from the analytical expression of eq \ref{t-deltac}. The good match between simulation data and calculated values establishes the self-consistency of the model.\\

\textbf{First crystallization time dependence on Diffusion coefficient}\\
Salt playas can have salts of different chemical compositions depending upon the geographical location. Principal salts in salt pans are halite ($NaCl$), but other salts like gypsum ($CaSO_{4}.2H_{2}0$),
mirabilite ($Na_{2}SO_{4}. 10H_{2}O$), thenardite ($Na_{2}SO_{4}$),
epsomite ($MgS0_{4}.7H_{2}O$) and trona
($NaHCO_{3}. Na_{2}CO_{3}.2H_{2}O$) are found too. 
Potash-rich saline pans have been reported
from the Chaidam Basin of China \cite{Qi1993, Wang2023} that are rich in sylvite ($KCl$) and
carnallite ($KCl.MgCl_{2}. 6H_{2}O$.  The diffusion coefficient for each of these salt solutions will be different under identical ambient conditions. As our model is generic, an analysis was done to check the dependence of salt crystallization time on the diffusion coefficient of the saline solution.  We observe that $t_{cr}$ increases non-linearly with increasing $D$, and displays a power-law dependence when plotted on a log-log scale with $\Delta c$ as depicted in Figure \ref{t-c}c. Thus
\begin{equation}
\log{t_{cr}} = m^{\prime}_{\Delta c}\log{D} +k^{\prime}_{\Delta c}
\label{m2}
\end{equation}

As $D$ increases, the time of first crystallization $t_{cr}$ increases as expected since diffusion impedes the attainment of supersaturation of the salt at the top layer of the channel for all $\Delta c$. 
Both $m^{\prime}_{\Delta c}$ and $k^{\prime}_{\Delta c}$ are functions of $\Delta c$.  $m^{\prime}_{\Delta c}$ increases exponentially with $\Delta c$ as displayed in the inset of Figure \ref{t-c}c, while  $k_{\Delta c}$ shows a parabolic variation with $\Delta c$.  
  \begin{eqnarray}
   m^{\prime}_{\Delta c} &=& \alpha^{\prime} e^{\beta^{\prime}\Delta c} \\
   k^{\prime}_{\Delta c} &=& 4a^{\prime} (\Delta c - \zeta)^{2} + k^{\prime}_0
  \label{m2-deltac}
  \end{eqnarray}
Here $\alpha^{\prime}$, $\beta^{\prime}$, $a^{\prime}$, $\zeta$ and $k^{\prime}_0$ are constants for constant evaporation flux. Of the factors $\alpha^{\prime}$, $\beta^{\prime}$ and $k_{0}^{\prime}$ are positive while $a^{\prime}$ is negative ; however for the range of $d$ and $p$ values chosen, the exponent remains positive. Combining eqs ( \ref{m2} -- \ref{m2-deltac}), we get --
\begin{equation}
 t_{cr}=  \exp\left[{\alpha^{\prime} e^{\beta^{\prime}\Delta c} \log D + 4a^{\prime} (\Delta c - \zeta)^{2} + k^{\prime}_0}\right]
\label{t-deltac-prime}
\end{equation}
Equation \ref{t-deltac-prime} is validated from the simulation results displayed in Figure \ref{t-c}d where the exponential curve  (denoted by lines) drawn using eq \ref{t-deltac-prime} almost exactly fits the simulation data obtained for $t_{cr}$ versus $\Delta c$ for constant values of $p$. 
Thus the simulation model produces self-consistent results.

When Figure \ref{t-c}c is examined carefully for higher values of $D$, demarcated by a vertical dashed line, it is observed that the first crystallization time $t_{f}$ becomes a constant for all $\Delta c$ for any constant $p$. This is more clearly evident for $\Delta c \geqslant 0.56$ in the figure. It is to be noted that while advection flow transfers mass in only the direction of the pressure field, diffusion distributes mass equally to all the neighbouring cells, i.e. in effect only $\frac{1}{6}^{th}$ of the mass transferred by diffusion is directed opposite to advection mass transfer. This is true for any value of $\Delta c$. When $\Delta c$ is high (in our simulation $\geqslant 0.56$), as the top layer of the channel starts crystallization, the combative contribution of the partial diffusive mass transfer in the direction opposite to advection is erased by the extremely small difference in salinity that exists between successive layers at the top of the channel, Figure \ref{fig:3D-pattern}b. Hence $t_{cr}$ shows no further change with an increase in $D$, i.e. it becomes diffusion-limited.\\

\textbf{Crystal growth}\\
\begin{figure}[H]
\centering
\includegraphics[width=0.85\linewidth]{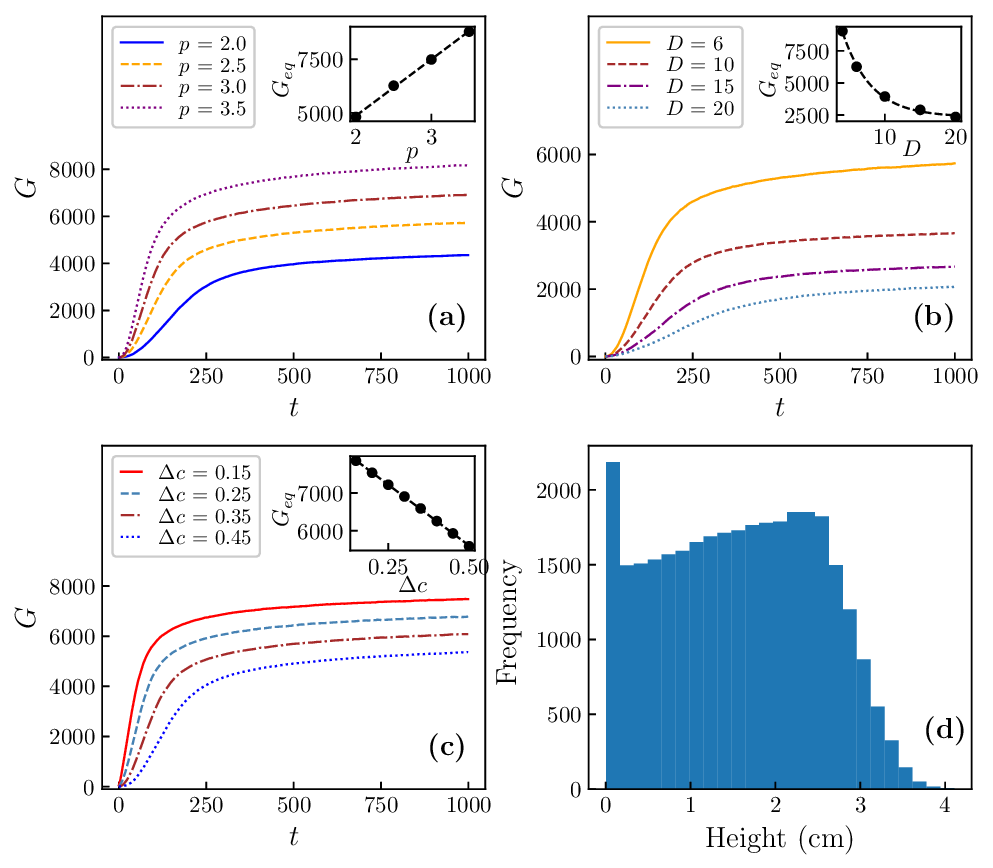}
\caption{All parameters are expressed in suitable units. (a-c) Effect of the model parameters $p, D$ and $\Delta c$ on crystal growth rate versus time, after crack maturation. The crystal growth rate $G$ shows a sigmoid function of $t$ and it reaches a dynamic equilibrium value $G_{eq}$ for each case. The variation of $G_{eq}$ with the variable parameter is displayed as inset to the figures. (a) Advection pressure $p$ variable with constant $D=6 $ and $\Delta c =0.5 $. (b) Diffusion coefficient $D$ variable with constant $p=2.5$ and $\Delta c=0.5$. (c)  $\Delta c$ variable with constant $p=2.5$ and $D=6$. (d) A histogram of a typical height distribution of salt ridges tessellating the playa. This histogram was generated with $p=2.5$, $D=6$, $\Delta c=0.4$ and after 1000 time steps. }
\label{growth}
\end{figure}
The effect of each of the variables $p$, $D$ and $\Delta c$ on crystal growth rate $G$ was tracked with time on the matured crack mosaic; when one parameter was varied, the other parameters were held constant. In each case, the salt growth rate increased sharply at first and then levelled to a constant plateau when the flow reached a state of dynamic equilibrium, showing a sigmoidal variation, Figures (\ref{growth}a - \ref{growth}c). The sigmoid function is a cumulative distribution function of the log-normal distribution.  The log-normal distribution has been reported specially in the case of random nucleation and growth processes\cite{Garside1983} where the crystallization seed is formed randomly during the simultaneous growth of other crystals. However, these growth processes have also been reported to occur otherwise\cite{Bergmann1997}.

The crystal growth rate increased sharply within approximately the first 100 time steps, though this variation in rate is different for the parameters $p$, $D$ and $\Delta c$; the rise was sharpest with change in $\Delta c$ and slowest for change in $D$. 
When $\Delta c$ is varied the flow control parameters $p$ and $D$ are kept constant; hence the sharp increase is attributed to the quick attainment of super-saturation in the top few layers of the pore channel that are quite close to $c_{cr}$. Figure \ref{fig:3D-pattern}b shows that below the top layers in a channel, the salinity diffuses out rather sharply. Hence the rate of crystallization flattens which is indicative of the interplay of $p$ and $D$ before dynamic equilibrium $G_{eq}$ is reached. The value of $G_{eq}$ decreases linearly with increasing $\Delta c$ as shown in the inset of Figure \ref{growth}c.

The insets of Figures \ref{growth}a and \ref{growth}b show that while $G_{eq}$ increases monotonically and linearly with an increase in advection pressure $p$; when $D$ is increased $G_{eq}$ decreases exponentially.  In any flow system, the diffusion velocity is the slowest and only $\frac{1}{6}^{th}$ of its value can oppose the advection flow direction in a 3-D system. Therefore as $D$ increases, the crystal growth rate change takes a longer time to decrease.

Figure \ref{growth}d displays a typical histogram of ridge height across the salt playa after 1000 time-steps. This data was collected using arbitrarily chosen values of $p=2.5$, $D=6$ and $\Delta c=0.4$. The ridge height is $\approx 2.8cm$ across the salt playa which is of the same order of observed reports\cite{Lowenstein1985}.  This indicates that the order of $p$, $D$, $\Delta c$ and the angle of repose are realistic.

\subsection*{System Energy and Voronoi-ness}
  The total energy $E$ of a mud plane with deep vertical cracks can be expressed as a sum of three terms -- (i) $E_0$ which factors all energy contributions except elastic energy and fracture energy, (ii) elastic energy of the mud columns and (iii) fracture energy spent to create new crack area. In our model, we assume $E_0$ to be constant for playas under similar ambient conditions. The evolution of the playas under identical conditions is determined by the mechanical property of the system. Therefore during the process of crack evolution, the total mechanical energy change in a simulation time-step involves contributions from the last two terms only.

    The elastic energy at any time step is proportional to the change in volume of the matrix columns in that time step.  This is given by $\gamma \sum_{i} A_{i} \delta z$, where $A_{i}$ is the cross-sectional area of the $i^{th}$ column and $\delta z$ is the change in the depth of the column in the time step.  We assume that the playa matrix is homogeneous and isotropic in the plane transverse to the direction of evaporation. Hence all matrix columns crack by equal infinitesimal vertical depths of $\delta z$ in a single time step. $\gamma$ is a parameter that represents the characteristics of the playa system. During each wetting-drying cycle, a new surface of $L \delta z$ opens up, where $L$ is the total length of the crack channels in the transverse plane. The fracture energy to create this new surface is given by $\sigma L \delta z$, where $\sigma$ is the fracture energy per unit area. Therefore, the total energy of the system per unit height is
    \begin{equation}
    E = \gamma \sum_{i} A_{i} + \sigma L
    \label{energy}
    \end{equation}
 
  Plotting the total area of the polygons and the length $L$ at every time step reveals that the total area of the polygons is practically constant. However, it is the total crack length $L$ that changes with time. Thus, while the area of the playa remains constant in our simulation, the shape of the polygons changes with wetting-drying cycles. The shape of the polygons can be quantified using the iso-perimetric ratio defined as $\lambda = \frac{4\pi A}{L^{2}}$, where $A$ is the area of a polygon and $L$ is its perimeter. The value of $\lambda$ is 1 for circles and for needle-like polygons, the value approaches 0. For a mosaic, the iso-perimetric ratio is the average over all the polygons in the mosaic, $\lambda = \frac{1}{M}\sum_{i=1}^{M} \lambda_{i}$.  Substituting $L$ in terms of $\lambda$ and the total area of the playa as constant in eq \ref{energy},  the total mechanical energy at every time step is expressed in terms of the iso-perimetric ratio (for details see Haque et. al\cite{Ruhul2022, Ruhul2023chaos})
  \begin{equation}
      E = \rho \lambda^{-\kappa}
  \end{equation}
  where $\rho$ and $\kappa$ are constants. Figure \ref{energy}a shows that the energy decreases with wetting--drying cycles as the patterns become stable.
  
\begin{figure}[H]
  \centering
  \includegraphics[width=0.85\linewidth]{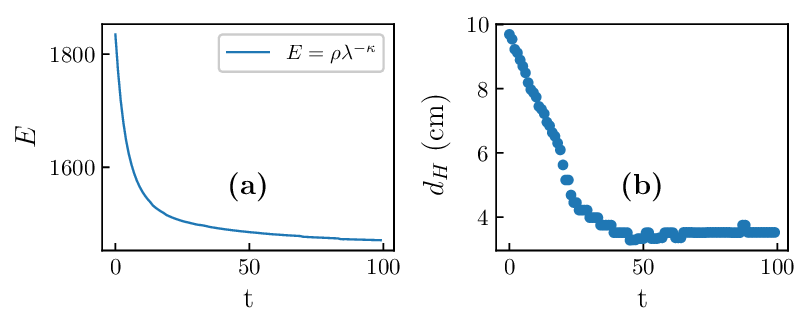}
  \caption{All parameters are expressed in suitable units. (a) Energy evolution of the system (b) Evolution of Hausdorff metric between the salt ridge pattern and the corresponding Voronoi diagram with wetting--drying cycles. }
  \label{energy}
\end{figure}

The Hausdorff metric for the salt ridge pattern and the corresponding Voronoi diagram is measured at each wetting--drying cycle as shown in Figure \ref{energy}b. The Hausdorff metric decreases with the wetting--drying cycles as the ridge pattern becomes more Voronoi-like. Thus the evolution of salt playas is an effort towards the minimization of total system energy during which the playa graduates towards a centroidal Voronoi mosaic.

\section{Conclusions}
Salt playas with their tessellated surface of hexagonal polygons of salt ridges are beautiful and intriguing, and have been a challenge to the scientific community for a realistic and physically meaningful model that can provide a thorough explanation of their formation. In this work, we have investigated the formation phenomenon via suitable modelling and simulation of the dynamical processes that are responsible.

Principles of fracture mechanics have been used to simulate crack formation of the playa surface using a spring network model. Periodic wetting-drying cycles are mimicked by their effects on the spring model that allow the initial crack mosaic tessellating the playa surface to mature into almost hexagonal polygons that tile the playa to minimize the net energy of the system. Surface cracks proceed vertically downwards with time to form crack conduits connecting the subsurface water table to the playa surface.

 We implement fluid transport in the crack conduits to establish the pressure and velocity field in the channels by solving the Navier-Stokes equation following the finite difference method.  The evaporation flux determines the pressure on the playa surface that draws the salty water from the subsurface water table to the surface. Evaporation from the top of the channels is a stochastic process that results in an increase in salt concentration there. The salinity distribution in a typical crack channel is determined by solving the advection-diffusion equation of mass transport. Salt crystallization has been allowed on the crack edge of a channel provided a critical salt concentration is reached at the top layer. 
It is assumed that the crack edge roughness presents nucleation sites for crystallization to occur. 

As the deposited salt at the crack edge grows in time, the ridge self-organises to distribute the heaped salt to neighbouring sites with equal probability to maintain an angle of repose as reported in the literature. However, we have not allowed salt melting in our model. The parameters of the model - evaporation flux pressure $p$, salt diffusion coefficient $D$, the difference between the initial and supersaturation salinity $\Delta c$ have been examined separately to investigate the time of first crystallization $t_{cr}$ and growth rate dynamics. $t_{cr}$ is a function of all three parameters in general, but is a function of $p$ and $\Delta c$ for any given salt playa. We have analysed the interdependence of the parameters on $t_{cr}$  and found the results consistent with our simulation results.

 The growth rate of the salt ridges reaches a state of dynamic equilibrium eventually. The rate of change of crystal growth rate follows a log-normal dependence with time as reported in random nucleation during crystallization growth processes. The diffusion coefficient of different salt playas appears to be the most important parameter that determines the attainment of dynamic equilibrium,  establishing mass transport to be diffusion-limited.

Combining all the various dynamical processes judiciously, we have successfully modelled the Voronoi-like mosaic of the salt ridges that tessellate salt playas as evident from real images of the same. Further, It has been proved that all these non-linear processes occur in tandem to minimize the system energy.

We have employed fracture mechanics, principles of energy minimization of system, fluid and mass transport in fracture channels and processes of crystallization and self organisation to finally replicate the almost Voronoidal pattern of salt ridges that tessellate salt playas. To the best of our knowledge, this is the first complete exposition of the phenomena -  Voronoidal pattern of ridges of salt observed in salt playas.


\section{Acknowledgement}
R A I Haque acknowledges support from UGC-funded research fellowships (UGC-ref No. 1435/ CSIR-UGC NET JUNE 2017). The authors are grateful to Prof. S. Tarafdar for useful discussions. The authors would like to thank  Richard Gibson and Michael Stickney for their helpful comments on the geology of salt playas.

\bibliography{ref_sp}

\end{document}